\documentclass[usenatbib]{mnras}

\usepackage{newtxtext,newtxmath}

\usepackage[T1]{fontenc}
\usepackage{ae,aecompl}

\usepackage{graphicx}	
\usepackage{amsmath}	
\usepackage{amssymb}	
\usepackage[version=4]{mhchem}

\newcommand{\Xstate}{\mathrm{X}\,{}^1\Sigma^+}
\newcommand{\Astate}{\mathrm{A}\,{}^1\Pi}
\newcommand{\astate}{\mathrm{a}\,{}^3\Pi}
\newcommand{\bstate}{\mathrm{b}\,{}^3\Sigma^-}
\newcommand{\cstate}{\mathrm{c}\,{}^3\Sigma^+}
\newcommand{\dstate}{\mathrm{d}\,{}^3\Pi}
\newcommand{\CHion}{\mathrm{CH}^+}
\newcommand{\CatomP}{\mathrm{C}({}^3\mathrm{P})}
\newcommand{\CionP}{\mathrm{C}^+({}^2\mathrm{P^o})}
\newcommand{\Cion}{\mathrm{C}^+}
\newcommand{\Hion}{\mathrm{H}^+}

\newcommand{\Hatom}{\mathrm{H}}

\newcommand{\rateunit}{\mathrm{cm}^3/\mathrm{s}}

\title[Radiative association of $\CatomP$ and $\Hion$]{Radiative association of $\CatomP$ and $\Hion$: Triplet States}

\author[J. F. Babb  and B. M. McLaughlin]{
                         James F. Babb$^{1}$\thanks{E-mail: jbabb@cfa.harvard.edu} and 
                         Brendan  M. McLaughlin$^{1,2}$\thanks{E-mail:bmclaughlin899@btinternet.com}
\\
$^{1}$Institute for Theoretical Atomic Molecular and Optical Physics (ITAMP), Harvard-Smithsonian Center for Astrophysics,\\ 
           60 Garden St., Cambridge, MA 02138, USA\\
$^{2}$Center for Theoretical Atomic Molecular and Optical Physics (CTAMOP),  School of Mathematics and Physics, \\
            The David Bates Building, 7 College Park,  Queens University Belfast, Belfast BT7 1NN, UK}

\date{Accepted XXX. Received YYY; in original form: \today}
\pubyear{2017}

\begin{document}
\label{firstpage}
\pagerange{\pageref{firstpage}--\pageref{lastpage}}
\maketitle


\begin{abstract}
The radiative association of $\CatomP$ and $\Hion$ is
investigated by calculating cross sections
for photon emission into bound ro-vibrational
states of $\CHion$ 
from the vibrational continua of initial triplet $\dstate$ or $\bstate$ states. 
Potential energy curves and transition dipole
moments are calculated using multi-reference configuration interaction (MRCI) methods
with AV6Z basis sets.
The 
cross sections are evaluated using quantum-mechanical methods and rate coefficients are calculated. 
The rate coefficients are about 100 times larger
than those for radiative association of 
$\CionP$ and $\Hatom$ from the $\Astate$ state.
We also confirm that the formation of $\CHion$ by radiative association of $\CionP$ and $\Hatom$ via the triplet $\cstate$ state is a minor process.
\end{abstract}


\begin{keywords}
ISM:molecules -- molecular processes -- astrochemistry -- molecular data -- scattering 
\end{keywords}


\section{Introduction}
The methylidyne ion $\CHion$ 
(methylidynium)
 is a
prominent interstellar molecular ion and the
mechanisms of its formation 
in various astrophysical environments continue to be of
interest~\citep{BlaDal73a,SteWil74,DalBla76a,TalDeF91,Wil92,Tie05,IndOkaGeb10,NagVanOss13,BacMonWie13,MorGupNag16}.
In particular, the formation of $\CHion$ by radiative association of $\Cion$ and $\Hatom$,
\begin{equation}
\label{RAsinglet}
\ce{\CionP{} + H -> CH+ + h\nu}
\end{equation}
was explored using collision theory some time ago
by \citet{Bat51} and
in subsequent improved calculations ~\citep{SolKle72,SmiLisLut73,GiuSuzRou76,GraMosRou83,BarvanHem06}.
No calculations, however, yielded a rate coefficient for the process (\ref{RAsinglet})  that
was large enough to explain the observed abundance of $\CHion$
in diffuse interstellar clouds~\citep{GiuSuzRou76,BarvanHem06}
and, depending on the application,
other formation mechanisms are now considered
of more significance~\citep{BlaDal73a,DalBla76a,BlaDal77,SteWil74,Wil92,CecPes10,NagVanOss13}.

\begin{figure}
                \includegraphics[width=\columnwidth]{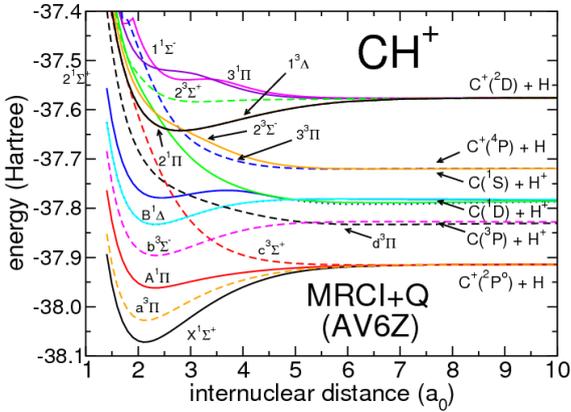}
    \caption{Potential energy curves for $\CHion$; singlet (solid lines) 
             and triplet molecular states (dashed lines) as 
             a function of internuclear distance.  All quantities are in atomic units (1 Hartree $=e^2/a_0\approx 27.2114~\mathrm{eV}$).
     }
    \label{fig1}
\end{figure}

Prior studies of the reaction (\ref{RAsinglet})  considered
transitions of the singlet symmetry molecular states $\Astate \rightarrow \Xstate$ 
for the radiative association process in
$\CionP$ and $\Hatom$ collisions, as 
the $\cstate \rightarrow \astate$ transitions are expected
to be insignificant~\citep{GiuSuzRou76}.
In Fig.~\ref{fig1} the potential energy curves of $\CHion$ correlating to $\CionP+\Hatom$ are shown.
The most recent calculations find that the rate coefficient for the radiative association process 
(\ref{RAsinglet}) is about $\sim 3\times 10^{-17}\mathrm{cm}^3/\mathrm{s}$ at 200~K~\citep{BarvanHem06},
only a factor of two smaller
than that of \citet{GraMosRou83}.

Another radiative association process
leading to $\CHion$, for which quantitative
data on the rate coefficients are unavailable, 
is the
radiative association of
$\CatomP$ atoms and protons
\begin{equation}
\label{RAtriplet}
\ce{C ({}^3P) + H+ -> CH+ + h\nu},
\end{equation}
which takes place via initial  molecular channels correlating to 
$\CatomP + \Hion$
and final channels yielding $\CHion$, namely
$\dstate\rightarrow\astate$, $\dstate\rightarrow\bstate$,
and $\bstate\rightarrow\astate$, as illustrated in Fig.~\ref{fig1}.
Inspection of Fig.~\ref{fig1} reveals that the triplet transition $\dstate\rightarrow\cstate$ is between
two repulsive molecular states and does not lead to significant bound $\CHion$.
\citet{StaGuHav98} assessed radiative collisions between
$\CatomP + \Hion$ leading to $\CionP + \Hatom$, but
detailed calculations for the radiative association
process (\ref{RAtriplet}) are not available.
In the present paper, we calculate, using quantum-mechanical collision theory,
the cross sections and rate coefficients for
the radiative association process (\ref{RAtriplet}) via 
the $\dstate\rightarrow\astate$, $\dstate\rightarrow\bstate$,
and $\bstate\rightarrow\astate$  transitions.  
For completeness, we also assess quantitatively the 
$\cstate\rightarrow\astate$ transitions and 
confirm their insignificance for radiative association.

\section{Theory}
\label{sec:theory}

The quantum mechanical theory of  radiative association  is well-established.
The cross section 
$\sigma_{i\rightarrow f}(E)$
for the process can be calculated using perturbation theory,
see for example \citep{StaBabDal93,Nyman2015},
\begin{equation}
 \sigma_{i\rightarrow f} (E)  = \sum_{v^{\prime}J^{\prime}}^{}\sum_{J}^{}
  \frac{64}{3} \frac{\pi^5\hbar^2}{c^3} \frac{\nu^3}{2\mu E} P_{i}S_{J J^{\prime}}
   |M_{i E J, f v^{\prime} J^{\prime}}|^2,
\label{eq:cross}
\end{equation}
where the sum is over the initial rotational $J$, final vibrational $v^{\prime}$, 
and rotational $J^{\prime}$ quantum numbers,
$i$ and $f$, respectively, denote the initial and final electronic states,
and $h\nu=E + |E_{v'J'}|$ is the photon energy.
The cross sections depend on the 
relative kinetic energy $E=\hbar^2k^2/2\mu$ of the $\CatomP$ and $\Hion$ 
(or in the case of the $\cstate$--$\astate$ states the $\CionP$ and $\Hatom$) colliders,
with momentum $\hbar k$,
where $\mu$ is the reduced mass.
The statistical weight factors $P_{i}$ are $\frac{2}{3}$ for the $\dstate$ state,
$\frac{1}{3}$ for $\bstate$ state, and $\frac{1}{4}$ for $\cstate$ state,
$S_{J,J^{\prime}}$ are the appropriate line strengths \citep{Cowan1981,Curtis2003} 
or H\"{o}nl-London factors \citep{Watson2008}, and $c$ is the speed of light.  
The matrix element $M_{i E J, f v^{\prime} J^{\prime}}$ is given by the integral
\begin{equation}
{ M_{i E J, fv^{\prime} J^{\prime}}}
=\int_{0}^{\infty} F_{i E J}(R) D(R) 
\Phi_{f v^{\prime} J^{\prime}} (R)dR,
\label{matrix}
\end{equation} 
where
the wave function  $\Phi_{f v^{\prime} J^{\prime}} (R)$
is a bound state eigenfunction on the final electronic state,
$D(R)$ is the transition dipole moment function, and 
$F_{i EJ} (R)$ is an energy-normalized continuum wave function
on the initial electronic state.
The bound and continuum state wave functions may be computed from their
respective Schr\"odinger equations \citep{MottMassey1965} 
using standard methods~\citep{Cooley1961,Johnson1977}. 

Rate coefficients are calculated by averaging the cross section over a
Maxwellian velocity distribution and are given by
 \begin{equation}
 \label{thermal-rate}
 \alpha (T) = \left( \frac{8}{\pi\mu } \right)^{1/2} \left( \frac{1}{k_B T} \right)^{3/2} \int_{0}^{\infty} E ~ \sigma_{i \rightarrow f} (E) \exp \left( - \frac{E}{k_B T} \right) d E,
 \end{equation}
where $k_B$ is the Boltzmann constant.

\section{Calculations}\label{sec:calculations}
\subsection{Molecular Structure}\label{sec:structure}
The low-lying potential energy curves of $\CHion$, $\Xstate$ and $\Astate$, 
have been extensively studied,
see, for example,  \citet{ElaOddBee77,SaxKirLiu80}
and the summary in \citet{ChoLeR16}.
Detailed calculations have also been performed for the excited 
 $\astate$ and $\cstate$ states  by
\cite{GreBagLiu72,GraMosRou83,BarvanHem04,OCoBecBla16}. 

Extended calculations provide information on the 
$\bstate$ state~\citep{SaxLiu83} and the $\dstate$ state correlating to 
$\CatomP  + \Hion$~\citep{LevRidLeC85,StaGuHav98,BigShaMag14}.
Other recent sophisticated calculations include the quasi-degenerate
many body perturbation theory (MBPT) approach of \citet{Freed1991}, 
the  equation of motion coupled 
cluster (EOMCCSDT) method developed by \citet{KowPie01}, the
multi-reference perturbation theory (MRPT) used by \citet{SelKhr12,SelKhrSte13},
and the effective (mass-dependent) calculations of \citet{SauSpi13}.

In order to perform dynamical calculations for the 
radiative association cross sections and rates for process (\ref{RAtriplet}), we require the 
potential energy curves (PECs) for the triplet states
of the entrance and exit channels, and the 
transition dipole  moment (TDM) coupling the states.

Some of the necessary data are available in the literature mentioned above. In particular, \citet{BigShaMag14} 
used the quantum chemistry suites \textsc{orca}
and \textsc{gamess-us} in multi-reference configuration interaction (MRCI) calculations with aug-cc-pV5Z (AV5Z) basis sets. 
We chose to calculate the potential energy curves and transition dipole moments using the MRCI approach 
with the \textsc{molpro} quantum chemistry suite,
using AV6Z basis sets.
In \textsc{molpro}, we use C$_{2\mathrm v}$ symmetry with the order of Abelian
irreducible representations being ($A_1$, $B_1$, $B_2$,  $A_2$). 
In reducing the symmetry from C$_{\infty\mathrm{v}}$ to 
C$_{2\mathrm v}$, the correlating relationships
are $\sigma \rightarrow a_1$, $\pi \rightarrow$ ($b_1$, $b_2$), and 
$\delta \rightarrow$ ($a_1$, $a_2$). 
We employed the non-relativistic 
state-averaged complete-active-space-self-consistent-field
 (\textsc{SA-CASSCF})/\textsc{MRCI} method~\citep{WerKno85,KnoWer85},
available within the \textsc{molpro} quantum chemistry 
codes \citep{MOLPRO_brief},
to take account of short-range interactions.
In detail, for this cation, six molecular orbitals
(MOs) are put into the active space, including four $a_1$, one $b_1$ and
one $ b_2$ symmetry MO's with all electron active.
The  molecular orbitals for the \textsc{MRCI} procedure were obtained from the 
state-averaged-multi-configuration-self-consistent-field 
\textsc{(SA-MCSCF)} method. In our work the Davidson 
correction (+Q) \citep{Davidson1974} 
was also applied to all molecular states.
The averaging processes 
is carried out for the lowest five ($^1A_1$), 
five ($^1B_1$), five ($^1A_2$), five ($^3A_1$), five  ($^3B_1$), 
and five ($^3A_2$) molecular states of 
the molecular ion in C$_{2\mathrm v}$.

\begin{table}
\centering
\caption{    Equilibrium bond distance $R_e$  (\AA{}) and
             dissociation energies $D_e$ (eV) for the $\Xstate$,  $\astate$, 
             $\Astate$ and $\bstate$ states of $\CHion$ for
             the present \textsc{MRCI+Q} calculations compared to other 
             theoretical and experimental results. 
             (The data are given in units conventional to
             quantum chemistry with 1~\AA{}=$10^{-10}$~m and $0.529177$~\AA{} $\approx a_0$.)}
\label{tab1}
\begin{tabular}{llll}
 \hline \noalign{\vskip 1mm}  
  State 	& Method  		&$R_e/$\AA{} 		& $D_e/\mathrm{eV}$\\
\hline
\noalign{\vskip 2mm}
$\Xstate$	&	&					&\\
		   	&MRCI+Q$^a$		&1.1256          		&4.291\\
		  	&MRCISD$^b$		&1.130	            	&4.244\\
			&MCSCF+CI$^c$	&1.1290			&4.140\\
		  	&QD-MBPT$^d$   	&1.1250			&4.66\\ 
		  	&MRD-CI$^e$		&1.129			&4.01 \\	
			&EOMCCSDt$^f$	&1.1404			& -- \\
			&Experiment$^g$	&1.1308843(30)	&4.77(43)\\ 
			&Empirical$^h$		&1.1284625(58)	&4.26044(4)\\
\\
  $\astate$	&				&					&\\
			&MRCI+Q$^a$    	&1.1305	            	&3.0945\\
		        &MRCISD$^b$  	&1.135       		&3.0404\\
		      	&QD-MBPT$^d$ 	&1.1473			&3.44\\ 
			&MRD-CI$^e$		&1.135			&2.88\\
           		&EOMCCSDT$^f$	&1.1261			&3.4404\\
            		&Experiment$^i$	&1.1348		  	& -- \\
            \\
 $\Astate$	&				&					&\\
			&MRCI+Q$^a$    	&1.2264		      	&1.2980\\
	        		&MRCISD$^b$  	&1.2390  			&1.239\\
            		&MCSF+CI$^c$	&1.2600    		&0.9360\\
	      		&QD-MBPT$^d$ 	&1.2055			&1.53\\ 
			&MRD-CI$^e$		&1.243			&1.00\\
            		&Experiment$^i$	& --				&1.2785\\  
              		&Empirical$^h$		&1.235896(14)		& 1.27750(3) \\
\\
 $\bstate$	&				&					&\\
			&MRCI+Q$^a$ 		&1.2373	  	       	&1.855\\
			&MRCISD$^b$    	&1.244	       		&1.833\\
			&MCSCF+CI$^c$	&1.2452			&1.854\\
			&MRD-CI$^e$		&1.245			&1.80\\	
            		&Experiment$^i$	&1.2416			& --\\
\hline
\end{tabular}
\begin{flushleft}
$^a$Multi-reference configuration interaction (MRCI) and Davidson correction (+Q), aug-cc-pV6Z basis, present work\\
$^b$MRCI with all single (S) and double (D) excitations, (MRCISD), aug-cc-pV5Z basis \citep{BigShaMag14}\\ 
$^c$Multi-configuration (MC) self-consistent field (SCF) + CI  \citep{SaxKirLiu80,SaxLiu83}\\
$^d$Quasi-Degenerate (QD) many body perturbation theory (MBPT)  \citep{Freed1991}\\
$^e$Multi-reference (MR) single and double-excitation CI (MRD-CI)  \citep{StaGuHav98}\\
$^f$Equation of motion (EOM) coupled cluster (CC) with S, D and triple (T) excitations  \citep{KowPie01}\\
$^g$Experiment, emission spectroscopy~\citep{Hakalla06}\\
$^h$Empirical fit of spectroscopic, photodissociation, and translational spectroscopy data  \citep{ChoLeR16} with energies
converted from $\mathrm{cm}^{-1}$ using the factor $1.239842\times 10^{-4}$~eV = 1 $\mathrm{cm}^{-1}$.\\
$^i$Experiment, photodissociation spectroscopy of stored ions \citep{HecRosLan07}\\
\end{flushleft}
\end{table}

These MOs ($4a_1$, $1b_1$, $1b_2$, $0a_2$), 
denoted by (4,1,1,0), were  generated from the state-averaging \textsc{CASSCF} process, 
and used to perform all the subsequent PEC calculations for all the 
electronic states in the \textsc{MRCI+Q} approximation.  Fig.~\ref{fig1} 
shows the calculated PECs for several low lying singlet and triplet states of 
the $\CHion$ molecular ion as a function of internuclear distance. 

Table \ref{tab1} gives a  comparison of the equilibrium bond length $R_e$, in \AA{},
and the dissociation energy $D_e$, in eV, determined from our  \textsc{MRCI+Q}  
work for a sample of low-lying states, namely;  
the $\Xstate$ ground state and  the excited 
$\astate$, $\Astate$, and $\bstate$  states, with the recent MRCI  
work of ~\citet{BigShaMag14} obtained using an AV5Z basis 
and with other previous work. 
Comparison of the present Table~\ref{tab1} with the calculations summarized in Table~2 of \citet{StaGuHav98} confirms 
that there has been substantial progress 
in the calculations for these electronic states of $\CHion$ in the last two decades.
We include the much more precise empirical parameters $R_e$ and $D_e$ for the 
$\Xstate$ and $\Astate$ states from the work of \citet{ChoLeR16}, 
who analyzed all available spectroscopic, photo-association,
and translational spectroscopy data, and who provide
a more extensive summary of experimental and theoretical results for these states.
Generally, we find that the present calculations
yield larger values of $D_e$ and smaller values of $R_e$ than those from
\citet{BigShaMag14}.

\begin{figure}
	\includegraphics[width=\columnwidth]{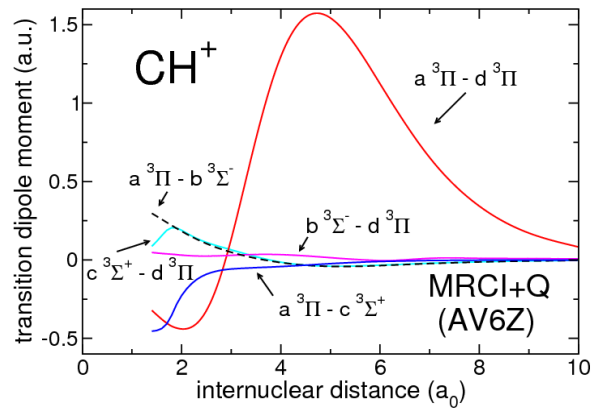}
    \caption{Transition dipole moments for the $\CHion$ cation coupling the 
             four low lying triplet molecular states, used in the present 
             radiative association cross section calculations,
             as a function of internuclear distance. All quantities are
             in atomic units.}
    \label{fig2}
\end{figure}

Our calculated TDMs are shown in Fig.~\ref{fig2} for the triplet transitions of interest
and they are in good agreement with those calculated by others using the \textsc{MRCISD} method~\citep{StaGuHav98,BigShaMag14},
where available.
The TDM connecting the $\bstate$ state to the $\dstate$ state
is not available, evidently, in the literature
and 
we list our calculated
values in Table~\ref{tab:d-b-tdm}.

As can be seen from Table~\ref{tab1} 
and Figs.~\ref{fig1}--\ref{fig2} our results are in good agreement with other multi-reference CI studies 
providing further confidence in our molecular data for the dynamical calculations.

%
\begin{table}
	\centering
	\caption{Transition dipole moment function $D(R)$ between the $\bstate$ state and $\dstate$ state in atomic units.
	The full table, for $1.4 \leq R \leq 10$, is available at \url{http://dx.doi.org/10.6084/m9.figshare.4725565}.
	%
	}
	\label{tab:d-b-tdm}
	\begin{tabular}{lr} 
    \hline
\multicolumn{1}{l}{$R/a_0$} & \multicolumn{1}{c}{$D(R)/e^2a_0^2$} \\
    \hline
1.4  & 0.047740 \\
1.5  & 0.044145 \\
1.6  & 0.041119 \\
1.7  & 0.038630 \\
1.8  & 0.036429 \\
1.9  & 0.033665 \\
2.0  & 0.030672 \\
\hline
	\end{tabular}
\end{table}
%

\subsection{Cross sections}
\label{sec:cross} 
\begin{figure}
                \includegraphics[width=\columnwidth]{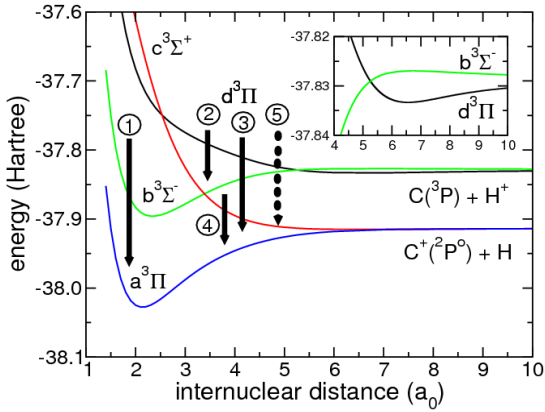}
    \caption{Potential energy curves for the four low-lying  triplet molecular 
states of the $\CHion$ molecular ion as a function of internuclear distance. 
The downward pointing solid arrows schematically indicate the transitions studied
in the present work for radiative association from: 1) the vibrational continuum of the $\bstate$ state to a bound ro-vibrational level of the $\astate$ state,
and similarly for, 2) $\dstate$ to $\bstate$, 3) $\dstate$ to $\astate$, and 4) $\cstate$ to $\astate$ transitions,
while, for 5) 
the dotted line connecting the vibrational continuum of
the $\dstate$ state to the vibrational
continuum of the $\cstate$ state indicates
that radiative charge transfer is the dominant mechanism for this channel.
The inset shows the potential energies of the $\bstate$ state and $\dstate$ state 
at higher energy resolution illustrating the barrier in the $\bstate$ state and
the shallow well of the $\dstate$ state. 
All quantities are in atomic units.
     }
    \label{fig3}
\end{figure}

In the present work, we require only the four low-lying 
potential energy curves (PECs) for the triplet $\astate$,
$\cstate$, $\bstate$, and $\dstate$ molecular states of Fig.~\ref{fig1}, which 
are plotted in Fig.~\ref{fig3} as a function of internuclear distance.
The necessary TDMs for the transitions 
are illustrated in Fig.~\ref{fig2}.  
It is apparent that the dominant contribution
to the  radiative association 
process (\ref{RAtriplet}) will come from 
the  $\dstate\rightarrow\astate$ transition, which has the largest transition dipole moment.
Moreover, at initial thermal kinetic energies, the vertical Franck-Condon overlap
and relatively small TDMs indicate that the 
$\cstate\rightarrow\astate$ and $\dstate\rightarrow\bstate$ transitions
will yield relatively small cross sections compared to the $\dstate\rightarrow\astate$ transition. 

We evaluated equation~(\ref{eq:cross}) using the calculated potential energies and TDMs of Sec.~\ref{sec:structure}. For values of $R<1.4\;a_0$, we joined the calculated potential energies to the corresponding values for $R=1.0$ and $1.2$, where available, from \citet{BigShaMag14}.
For $R >$ 10 $\mathrm{a}_0$, the appropriate long-range forms were used for the separating atom-ion pair.
In particular, for $\CatomP + \Hion$, this corresponds to a $R^{-3}$ quadrupole interaction added to the attractive $R^{-4}$ polarisation potential~\citep{GenGie77,LevRidLeC85}.
For $\CionP + \Hatom$, we used the the form
$-\frac{1}{2}\alpha_{\mathrm{H}}R^{-4}$,
where $\alpha_{\mathrm{H}}=\frac{9}{2}$ is the
static electric dipole polarisability of hydrogen. The potential energies for $R<10$ were
adjusted to match the long-range forms at $R=10$, with the asymptotic energies taken from
Table I, column 4, of \citet{StaGuHav98}.
The TDMs were fit to inverse powers of the internuclear distance for $R>10$
and extended to $R=1.0$ and 1.2 using the values, where available,
from \citet{BigShaMag14}.

\subsubsection{Triplet states of $\CatomP$ and $\Hion$}

\begin{figure}
	\includegraphics[width=\columnwidth]{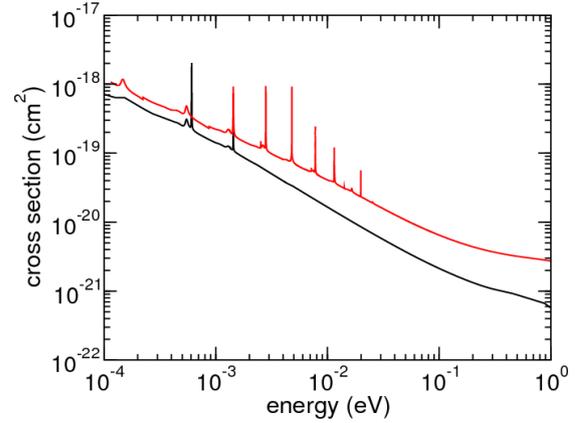}
    \caption{Cross sections (cm$^2$) for radiative decay (upper, red line) 
             and radiative association (lower, black line) as a function of the collision
             energy (eV), for the $\dstate$ $\rightarrow$ $\astate$ transition. }
    \label{fig:cross-d-a}
\end{figure}

Our calculated $\dstate$  state is attractive at long-range ~\citep{LevRidLeC85,SarWhi88} with a shallow
well at about 6.56~$a_0$
(see inset, Fig.~\ref{fig3})
and the transition dipole moment is large.
The radiative association cross sections
for the $\dstate$ to $\astate$ transitions are shown in Fig.~\ref{fig:cross-d-a}.
%
	%
For comparison, we calculated the radiative decay cross sections using the distorted wave optical potential approach
as detailed by~\citet{BabMcL17}.
The present results for radiative decay generally agree with those of \citet{StaGuHav98}; being about a factor two larger 
than theirs at $10^{-4}$~eV,  though gradually becoming comparable for increasing energies, and being equal for energies above 0.1~eV.
The differences may result from the use of different molecular potential energy data.
From \citet{StaGuHav98} we can obtain
an independent estimate of the radiative association cross sections.
We  subtracted 
their radiative charge transfer (denoted full quantum or ``FQ'') cross sections
from their  radiative decay cross sections (denoted optical potential distorted wave or ``OPDW'') given by, respectively, the solid line and dotted line with points in their Fig.~8. 
\begin{figure}
	\includegraphics[width=\columnwidth]{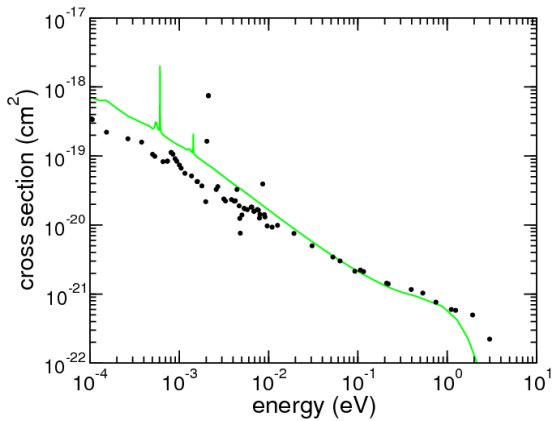}
    \caption{Cross sections for radiative
    association from the $\dstate$ state to the 
    $\astate$
    state for the present calculation (green line) and estimated (see text) from \citet{StaGuHav98} (black points).}
    \label{fig:compare-d-b}
\end{figure}
	%
In Fig.~\ref{fig:compare-d-b} we plot
the present cross sections and the estimate so obtained
and we find the agreement satisfactory for energies above
$10^{-2}$~eV. For lower energies, our values are about twice
as large. The discrepancy may arise from the reliability of the subtraction
procedure and the different molecular data used.

\begin{figure}
	\includegraphics[width=\columnwidth]{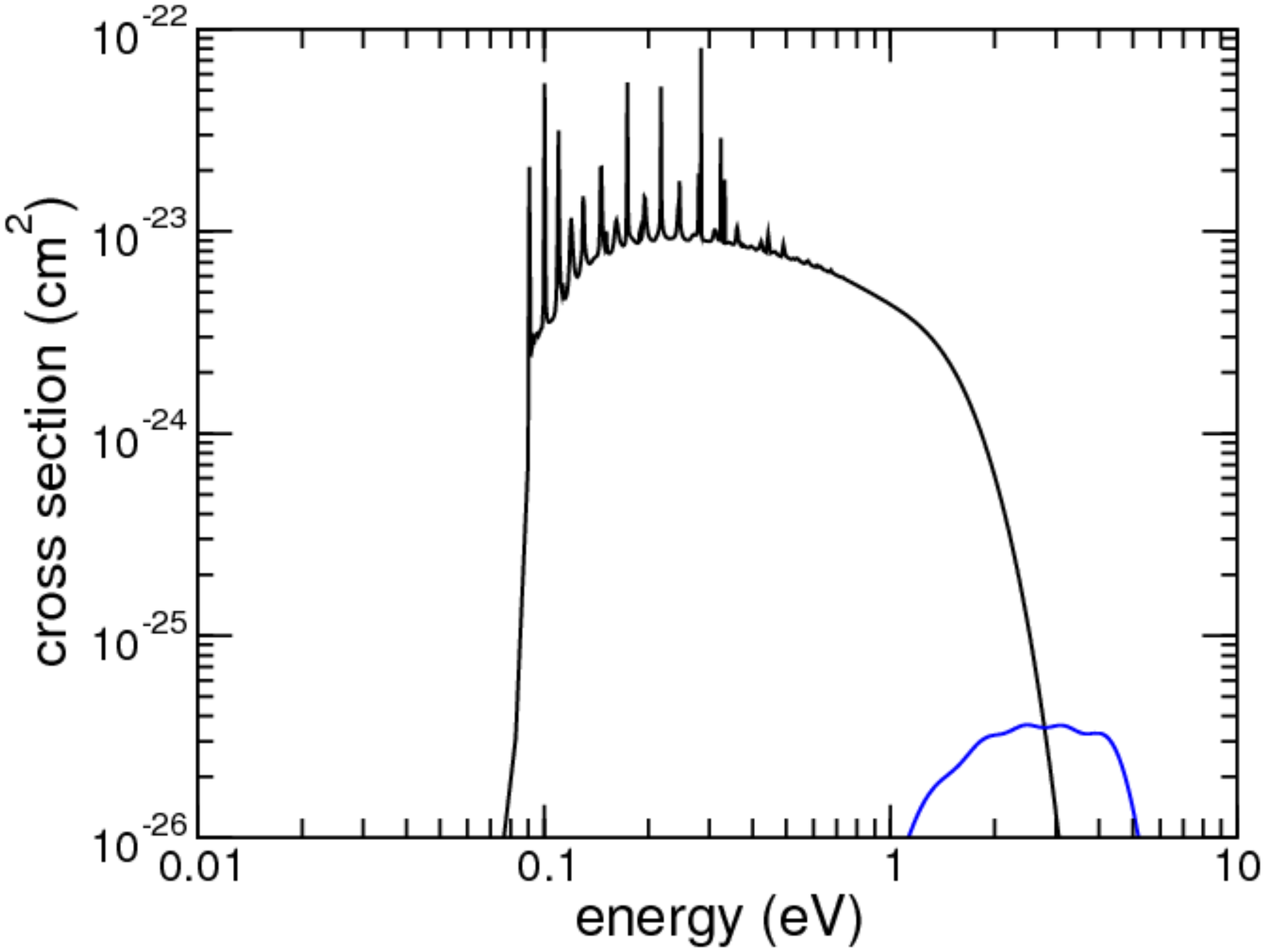}
    \caption{Cross sections (cm$^2$) as a function of collision energy (eV) 
             for radiative association from the $\bstate$ state to the $\astate$ state (black line)
             and from the $\dstate$ state to the $\bstate$ state (blue line). }
    \label{fig:cross-others}
\end{figure}

The cross sections for radiative association for the $\bstate$ to
$\astate$ transitions are shown in Fig.~\ref{fig:cross-others}.
	%
Because the $\bstate$ state is repulsive at long-range 
see inset, Fig.~\ref{fig3}),  
the cross sections drop off quite rapidly for energies lower than
the potential energy local maximum of $0.109$~eV (relative to
the $\CatomP + \Hion$ limit).
Numerous resonances are seen at higher energies;
however, as the cross sections are generally small, we did not
investigate the resonances in detail.
The radiative association cross sections are comparable to the  
radiative decay cross sections
for this channel given in Fig.~8 of \citet{StaGuHav98} but 
are, however, 
one hundredth of those for the 
$\dstate$ $\rightarrow$ $\astate$ transition.

We also considered the $\dstate\rightarrow\bstate$ transition and
the calculated cross sections, shown in Fig.~\ref{fig:cross-others}, are very small
due to limited Franck-Condon overlap.
Levels of the $\bstate$  state are predissociated
by the $\cstate$ state  for $v'\ga4$, see ~\citet{HecRosLan07}
for a detailed analysis, and the probability
of forming levels with $v'\la4$, which would
decay to the  $\astate$ state~\citep{BigShaMag14},
is small.
We ignored predissociation and radiative cascading and 
estimated the cross sections using the quantum-mechanical
approach, Eq.~(\ref{eq:cross}), but a
more detailed model including predissociation by the $\cstate$ state and
radiative decay from the $\bstate$ state to the $\astate$ state might be warranted.

There is negligible radiative
association between the $\dstate$ state and $\cstate$ state because
both states are repulsive with shallow long-range wells and
the transition dipole moment is small.
The cross section can be no more than the 
radiative decay cross section,
for which \citet{StaGuHav98} find
about $10^{-21}~\mathrm{cm}^2$ at $10^{-4}$~eV
decreasing to less than $10^{-23}~\mathrm{cm}^2$ at 1~eV.

\subsubsection{Triplet states of $\CionP$ and $\Hatom$}
We calculated the cross sections for radiative association from the $\cstate$ to the $\astate$ state.
The values are much smaller compared to those
given in the previous section.
We find that $\sigma (E)$ is about $8\times 10^{-28}~\mbox{cm}^2$ at
$10^{-3}$~eV to  about $2\times 10^{-28}~\mbox{cm}^2$ at 0.01~eV.

\subsection{Rate coefficients}
As we have shown, the $\dstate\rightarrow\astate$ transitions dominate
in the radiative association process (\ref{RAtriplet}).
We calculated the rate coefficients for radiative association 
and
radiative decay using  Eq.~(\ref{thermal-rate}).
The results are given in Table~\ref{tab:rates}.
\begin{table}
	\centering
	\caption{Rate coefficients, in units of $\rateunit$, as
	a function of temperature $T$, in K, for $\dstate\rightarrow\astate$ radiative decay and radiative association.
        A value in parentheses is a power of ten that should multiply the precedent quantity,
        for example, $1.01(-14)$ represents $1.01\times10^{-14}$.}
	\label{tab:rates}
	\begin{tabular}{lcc} 
	    \hline
\multicolumn{1}{l}{ } & \multicolumn{1}{c}{Radiative} & \multicolumn{1}{c}{Radiative}\\
\multicolumn{1}{l}{$T(K)$} & \multicolumn{1}{c}{Decay} & \multicolumn{1}{c}{Association}\\
    \hline
     10 & 1.01($-$14) & 6.48($-$15)  \\ 
     20 & 9.30($-$15) & 5.31($-$15)  \\ 
     30 & 8.40($-$15) & 4.48($-$15)  \\ 
     40 & 7.81($-$15) & 3.90($-$15)  \\ 
     50 & 7.46($-$15) & 3.64($-$15)  \\ 
     70 & 6.92($-$15) & 3.15($-$15)  \\ 
    100 & 6.20($-$15) & 2.79($-$15)  \\ 
    200 & 5.05($-$15) & 2.06($-$15)  \\ 
    300 & 4.46($-$15) & 1.73($-$15)  \\ 
    500 & 3.93($-$15) & 1.42($-$15)  \\ 
    700 & 3.59($-$15) & 1.26($-$15)  \\ 
   1000 & 3.35($-$15) & 1.14($-$15)  \\ 
   2000 & 3.24($-$15) & 9.87($-$16)  \\ 
   3000 & 3.24($-$15) & 9.36($-$16)  \\ 
   5000 & 3.45($-$15) & 8.59($-$16)  \\ 
   7000 & 3.62($-$15) & 7.72($-$16)  \\ 
  10000 & 3.81($-$15) & 6.60($-$16)  \\ 

\hline
	\end{tabular}
\end{table}
%

\begin{figure}
	\includegraphics[width=\columnwidth]{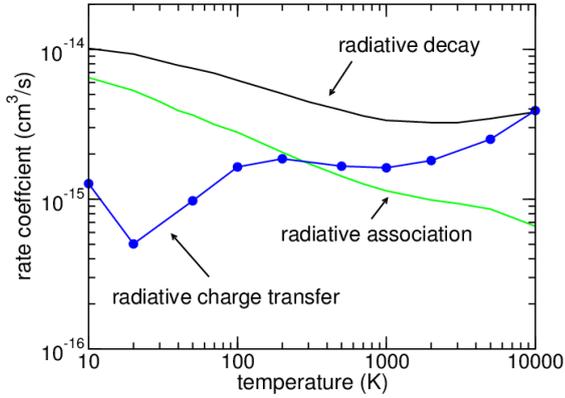}
    \caption{For the $\dstate\rightarrow\astate$ transition, the present rate coefficients for radiative association
    and radiative decay  
    compared to the rate coefficients from  \citet{StaHavKrs98} for radiative charge transfer (dots with guide line).
    }
    \label{fig:rates}
\end{figure}

The sum of the rate coefficients for the radiative association and the radiative charge transfer processes
should be approximately equal to the rate coefficients for radiative decay~\citep{CooKirDal84}.
As an additional
check of our results, we examined the temperature dependence of the present 
radiative association and radiative decay rate coefficients and
the radiative charge transfer rate coefficients
of \citet{StaHavKrs98} (who used using a different
set of molecular data), see Fig.~\ref{fig:rates}.
With increasing temperature, the radiative association rate coefficient diminishes 
and the radiative charge transfer coefficient increases, but their
sum remains close to the radiative decay rate coefficient.

We fit the calculated rate coefficients to the formula~\citep{NovBecBuh13,VisBuzMil16} 
\begin{equation}
\alpha(T) = a (300 / T)^x + T^{-3/2} \sum_{i=1}^3 c_i \exp (-d_i/T)\quad\rateunit \,,
\end{equation}
with $T$ expressed in K.
For radiative decay $(\dstate\rightarrow\astate)$, the values are
$a=4.89675\times10^{-15}$, $x=0.1797$, $c_1=7.35952\times10^{-9}$, $d_1=18087.6$,
$c_2=-3.38159\times 10^{-11}$, $d_2=758.448$,
$c_3=3.80018\times 10^{-13}$, and $d_3=23.763$.
Similarly, for the 
radiative association process ($\dstate\rightarrow\astate$), the values are 
$a=1.90925\times10^{-15}$, $x=0.3638$, $c_1=3.05111\times10^{-10}$, $d_1=7875.53$,
$c_2=-4.15024\times 10^{-12}$,  $d_2=427.166$, $c_3=0$, and $d_3=0$.
These parameters fit the data listed in Table~\ref{tab:rates} to better than ten percent across the
temperature range from 10 to 10000~K.

	%
	%
%
\section{Discussion}
We assessed the radiative association process for the  possible
triplet state transitions yielding $\CHion$.
The rate coefficients  will be dominated by the $\dstate$ to $\astate$ cross sections
and their values for radiative association (\ref{RAtriplet})
were calculated.
The rates for the $\dstate\rightarrow\astate$ transition 
decrease from about $ 6.5\times 10^{-15}\,\rateunit$ at 10~K to about 
$ 6.6\times 10^{-16}\,\rateunit$ at $10000$~K.

In our analysis, we ignored the effects of the spin-orbit splitting of the  $\CatomP$ atom and
the consequent fine structure splittings of the $\dstate$ state and $\bstate$ state~\citep{SarWhi88}
and their rotational coupling~\citep{StaGuHav98},
which may affect the validity of our results for low temperatures.
However, in the cold interstellar medium the $\CatomP$ atoms will be in the 
lowest level of the lowest term~\citep{Bat51},
which correlates with the $\dstate$ state~\citep{GenGie77}, making it likely
that the main channel for the radiative association  process (\ref{RAtriplet}) will be 
via the $\dstate\rightarrow\astate$ transition.
The chance that the $\CatomP$ atom and $\Hion$ ion come together in the $\dstate$ state,
which we assumed to be $\frac{2}{3}$, would need to be adjusted accordingly.

In most astrophysical environments
where formation of $\CHion$
is modeled, the carbon atoms are ionized before hydrogen
atoms, hence the process (\ref{RAsinglet}) is dominant, when it is
important,
where  the rate coefficients for the $\mathrm{A}{}^1\Pi\rightarrow\mathrm{X}{}~{}^1\Sigma^+$ transition are
around ${2}\times 10^{-17}\,\rateunit$ at 1000~K~\citep{GraMosRou83,BarvanHem06}.
Thus, while process (\ref{RAtriplet}) is faster than (\ref{RAsinglet}), its relevance will depend
on the relative concentration of C versus $\CionP$.

\section{Summary and Conclusions}

The rate coefficients for the radiative association process (\ref{RAtriplet})
were calculated and found
to be about 100 times larger than those for (\ref{RAsinglet}),
though applicability would require an environment where
neutral carbon exists in the presence
of protons, which is the reverse of the usual case.
An application might be to modeling the variation of the emission 
of atomic carbon with the cosmic ionization rate in metal-poor galaxies~\citep{GloCla16}.

\section*{Acknowledgements}
ITAMP is supported in part by a grant from the NSF to the Smithsonian
Astrophysical Observatory and Harvard University. 
BMMcL acknowledges support from the ITAMP visitor's program
and from Queen's University Belfast for the award 
of a Visiting Research Fellowship (VRF).   
Grants of computational time  at the National Energy Research  
Scientific Computing Center (NERSC) in Berkeley, 
CA, USA  and at the High Performance  Computing Center  Stuttgart (HLRS) of the 
University of Stuttgart, Stuttgart, Germany are gratefully acknowledged.
This research made use of the NASA Astrophysics Data System.



\bibliographystyle{mnras}

\begin{thebibliography}{}
\makeatletter
\relax
\def\mn@urlcharsother{\let\do\@makeother \do\$\do\&\do\#\do\^\do\_\do\%\do\~}
\def\mn@doi{\begingroup\mn@urlcharsother \@ifnextchar [ {\mn@doi@}
  {\mn@doi@[]}}
\def\mn@doi@[#1]#2{\def\@tempa{#1}\ifx\@tempa\@empty \href
  {http://dx.doi.org/#2} {doi:#2}\else \href {http://dx.doi.org/#2} {#1}\fi
  \endgroup}
\def\mn@eprint#1#2{\mn@eprint@#1:#2::\@nil}
\def\mn@eprint@arXiv#1{\href {http://arxiv.org/abs/#1} {{\tt arXiv:#1}}}
\def\mn@eprint@dblp#1{\href {http://dblp.uni-trier.de/rec/bibtex/#1.xml}
  {dblp:#1}}
\def\mn@eprint@#1:#2:#3:#4\@nil{\def\@tempa {#1}\def\@tempb {#2}\def\@tempc
  {#3}\ifx \@tempc \@empty \let \@tempc \@tempb \let \@tempb \@tempa \fi \ifx
  \@tempb \@empty \def\@tempb {arXiv}\fi \@ifundefined
  {mn@eprint@\@tempb}{\@tempb:\@tempc}{\expandafter \expandafter \csname
  mn@eprint@\@tempb\endcsname \expandafter{\@tempc}}}

\bibitem[\protect\citeauthoryear{Babb \& McLaughlin}{Babb \&
  McLaughlin}{2017}]{BabMcL17}
Babb J.~F.,  McLaughlin B.~M.,  2017, \mn@doi [{J. Phys. B: At. Mol. Opt.
  Phys.}] {10.1088/1361-6455/aa54f4}, 50, 044003

\bibitem[\protect\citeauthoryear{Bacchus-Montabonel \&
  Wiesenfeld}{Bacchus-Montabonel \& Wiesenfeld}{2013}]{BacMonWie13}
Bacchus-Montabonel M.-C.,  Wiesenfeld L.,  2013, \mn@doi [Chem. Phys. Lett.]
  {10.1016/j.cplett.2013.08.003}, 583, 23

\bibitem[\protect\citeauthoryear{Barinovs \& {van Hemert}}{Barinovs \& {van
  Hemert}}{2004}]{BarvanHem04}
Barinovs {\u{G}}.,  {van Hemert} M.~C.,  2004, \mn@doi [Chem. Phys. Lett.]
  {10.1016/j.cplett.2004.10.035}, 399, 406

\bibitem[\protect\citeauthoryear{Barinovs \& van Hemert}{Barinovs \& van
  Hemert}{2006}]{BarvanHem06}
Barinovs {\u{G}}.,  van Hemert M.~C.,  2006, \mn@doi [\apj] {10.1086/498080},
  636, 923

\bibitem[\protect\citeauthoryear{Bates}{Bates}{1951}]{Bat51}
Bates D.~R.,  1951, \mn@doi [MNRAS] {10.1093/mnras/111.3.303}, 111, 303

\bibitem[\protect\citeauthoryear{Biglari, Shayesteh  \& Maghari}{Biglari
  et~al.}{2014}]{BigShaMag14}
Biglari Z.,  Shayesteh A.,   Maghari A.,  2014, \mn@doi [Comp. Theo. Chem.]
  {10.1016/j.comptc.2014.08.012}, 1047, 22

\bibitem[\protect\citeauthoryear{{Black} \& {Dalgarno}}{{Black} \&
  {Dalgarno}}{1973}]{BlaDal73a}
{Black} J.~H.,  {Dalgarno} A.,  1973, \aplett, \href
  {http://adsabs.harvard.edu/abs/1973ApL....15...79B} {15, 79}

\bibitem[\protect\citeauthoryear{Black \& Dalgarno}{Black \&
  Dalgarno}{1977}]{BlaDal77}
Black J.~H.,  Dalgarno A.,  1977, \mn@doi [\apjs] {10.1086/190455}, 34, 405

\bibitem[\protect\citeauthoryear{Cecchi-Pestellini}{Cecchi-Pestellini}{2010}]{CecPes10}
Cecchi-Pestellini C.,  2010, in Babb J.~F.,  Kirby K.,   Sadeghpour H.,  eds,
  Proceedings of the {Dalgarno} Celebratory Symposium. Imperial College Press,
  London, p.~173

\bibitem[\protect\citeauthoryear{{Cho} \& {Le Roy}}{{Cho} \& {Le
  Roy}}{2016}]{ChoLeR16}
{Cho} Y.-S.,  {Le Roy} R.~J.,  2016, \mn@doi [\jcp] {10.1063/1.4939274}, \href
  {http://adsabs.harvard.edu/abs/2016JChPh.144b4311C} {144, 024311}

\bibitem[\protect\citeauthoryear{Cooley}{Cooley}{1961}]{Cooley1961}
Cooley J.~W.,  1961, \mn@doi [{Math. Comput.}] {10.2307/2003025}, \textbf{15},
  363

\bibitem[\protect\citeauthoryear{Cooper, Kirby  \& Dalgarno}{Cooper
  et~al.}{1984}]{CooKirDal84}
Cooper D.~L.,  Kirby K.,   Dalgarno A.,  1984, \mn@doi [Can. J. Phys.]
  {10.1139/p84-208}, 62, 1622

\bibitem[\protect\citeauthoryear{Cowan}{Cowan}{1981}]{Cowan1981}
Cowan R.~D.,  1981, {The Theory of Atomic Structure and Spectra}.
University of California Press, Berkeley, California, USA

\bibitem[\protect\citeauthoryear{Curtis}{Curtis}{2003}]{Curtis2003}
Curtis L.~J.,  2003, {Atomic Structure and Lifetimes: A Conceptual Approach}.
Cambridge University Press, Cambridge, UK

\bibitem[\protect\citeauthoryear{Dalgarno \& Black}{Dalgarno \&
  Black}{1976}]{DalBla76a}
Dalgarno A.,  Black J.~H.,  1976, \mn@doi [Rep. Prog. Phys.]
  {10.1088/0034-4885/39/6/002}, 39, 573

\bibitem[\protect\citeauthoryear{{Elander}, {Oddershede}  \& {Beebe}}{{Elander}
  et~al.}{1977}]{ElaOddBee77}
{Elander} N.,  {Oddershede} J.,   {Beebe} N.~H.~F.,  1977, \mn@doi [\apj]
  {10.1086/155457}, \href {http://adsabs.harvard.edu/abs/1977ApJ...216..165E}
  {216, 165}

\bibitem[\protect\citeauthoryear{Gentry \& Giese}{Gentry \&
  Giese}{1977}]{GenGie77}
Gentry W.~R.,  Giese C.~F.,  1977, \mn@doi [J. Chem. Phys.] {10.1063/1.435072},
  67, 2355

\bibitem[\protect\citeauthoryear{{Giusti-Suzor}, {Roueff}  \& {van
  Regemorter}}{{Giusti-Suzor} et~al.}{1976}]{GiuSuzRou76}
{Giusti-Suzor} A.,  {Roueff} E.,   {van Regemorter} H.,  1976, \mn@doi [J.
  Phys. B: At. Mol. Opt. Phys.] {10.1088/0022-3700/9/6/024}, \href
  {http://adsabs.harvard.edu/abs/1976JPhB....9.1021G} {9, 1021}

\bibitem[\protect\citeauthoryear{Glover \& Clark}{Glover \&
  Clark}{2016}]{GloCla16}
Glover S. C.~O.,  Clark P.~C.,  2016, \mn@doi [MNRAS] {10.1093/mnras/stv2863},
  456, 3596

\bibitem[\protect\citeauthoryear{Graff, Moseley  \& Roueff}{Graff
  et~al.}{1983}]{GraMosRou83}
Graff M.~M.,  Moseley J.~T.,   Roueff E.,  1983, \mn@doi [\apj]
  {10.1086/161088}, \href {http://adsabs.harvard.edu/abs/1983ApJ...269..796G}
  {269, 796}

\bibitem[\protect\citeauthoryear{{Green}, {Bagus}, {Liu}, {McLean}  \&
  {Yoshimine}}{{Green} et~al.}{1972}]{GreBagLiu72}
{Green} S.,  {Bagus} P.~S.,  {Liu} B.,  {McLean} A.~D.,   {Yoshimine} M.,
  1972, \mn@doi [\pra] {10.1103/PhysRevA.5.1614}, \href
  {http://adsabs.harvard.edu/abs/1972PhRvA...5.1614G} {5, 1614}

\bibitem[\protect\citeauthoryear{Hakalla, Kepa, Szajna  \& Zachwieja}{Hakalla
  et~al.}{2006}]{Hakalla06}
Hakalla R.,  Kepa R.,  Szajna W.,   Zachwieja M.,  2006, \mn@doi [{Eur. Phys.
  J. D.}] {10.1140/epjd/e2006-00063-9}, 38, 481

\bibitem[\protect\citeauthoryear{Hechtfischer, Rostas, Lange, Linkemann,
  Schwalm, Wester, Wolf  \& Zajfman}{Hechtfischer et~al.}{2007}]{HecRosLan07}
Hechtfischer U.,  Rostas J.,  Lange M.,  Linkemann J.,  Schwalm D.,  Wester R.,
   Wolf A.,   Zajfman D.,  2007, \mn@doi [{J. Chem. Phys.}]
  {10.1063/1.2800004}, 127, 204304

\bibitem[\protect\citeauthoryear{{Indriolo}, {Oka}, {Geballe}  \&
  {McCall}}{{Indriolo} et~al.}{2010}]{IndOkaGeb10}
{Indriolo} N.,  {Oka} T.,  {Geballe} T.~R.,   {McCall} B.~J.,  2010, \mn@doi
  [\apj] {10.1088/0004-637X/711/2/1338}, \href
  {http://adsabs.harvard.edu/abs/2010ApJ...711.1338I} {711, 1338}

\bibitem[\protect\citeauthoryear{Johnson}{Johnson}{1977}]{Johnson1977}
Johnson B.~R.,  1977, \mn@doi [{J. Chem. Phys.}] {10.1063/1.435384},
  \textbf{67}, 4086

\bibitem[\protect\citeauthoryear{Kanzler, Sun  \& Freed}{Kanzler
  et~al.}{1991}]{Freed1991}
Kanzler A.~W.,  Sun H.,   Freed K.~F.,  1991, \mn@doi [Int. J. Quant. Chem.]
  {10.1002/qua.560390306}, 29, 269

\bibitem[\protect\citeauthoryear{Knowles \& Werner}{Knowles \&
  Werner}{1985}]{KnoWer85}
Knowles P.~J.,  Werner H.-J.,  1985, \mn@doi [Chem. Phys. Lett.]
  {10.1016/0009-2614(85)80025-7}, 115, 259

\bibitem[\protect\citeauthoryear{Kowalski \& Piecuch}{Kowalski \&
  Piecuch}{2001}]{KowPie01}
Kowalski K.,  Piecuch P.,  2001, \mn@doi [Chem. Phys. Lett.]
  {10.1016/S0009-2614(01)01010-7}, 347, 237

\bibitem[\protect\citeauthoryear{Langhoff \& Davidson}{Langhoff \&
  Davidson}{1974}]{Davidson1974}
Langhoff S.~R.,  Davidson E.~R.,  1974, \mn@doi [Int. J. Quantum Chem.]
  {10.1002/qua.560080106}, 8, 61

\bibitem[\protect\citeauthoryear{Levy, Ridard  \& Le~Coarer}{Levy
  et~al.}{1985}]{LevRidLeC85}
Levy B.,  Ridard J.,   Le~Coarer F.,  1985, \mn@doi [Chem. Phys.]
  {10.1016/0301-0104(85)85024-2}, 92, 295

\bibitem[\protect\citeauthoryear{Morris et~al.,}{Morris
  et~al.}{2016}]{MorGupNag16}
Morris P.~W.,  et~al., 2016, \mn@doi [\apj] {10.3847/0004-637x/829/1/15}, 829,
  15

\bibitem[\protect\citeauthoryear{Mott \& Massey}{Mott \&
  Massey}{1965}]{MottMassey1965}
Mott N.~F.,  Massey H. S.~W.,  1965, The Theory of Atomic Collisions, 3rd edn.
Clarendon Press, Oxford, UK

\bibitem[\protect\citeauthoryear{{Nagy} et~al.,}{{Nagy}
  et~al.}{2013}]{NagVanOss13}
{Nagy} Z.,  et~al., 2013, \mn@doi [\aap] {10.1051/0004-6361/201220519}, \href
  {http://adsabs.harvard.edu/abs/2013A%26A...550A..96N} {550, A96}

\bibitem[\protect\citeauthoryear{Novotn{\'y} et~al.,}{Novotn{\'y}
  et~al.}{2013}]{NovBecBuh13}
Novotn{\'y} O.,  et~al., 2013, \mn@doi [\apj] {10.1088/0004-637x/777/1/54},
  777, 54

\bibitem[\protect\citeauthoryear{Nyman, Gustafsson  \& Antipov}{Nyman
  et~al.}{2015}]{Nyman2015}
Nyman G.,  Gustafsson M.,   Antipov S.~V.,  2015, \mn@doi [Int. Rev. Phys.
  Chem.] {10.1080/0144235X.2015.1072365}, 34, 385

\bibitem[\protect\citeauthoryear{{O'Connor} et~al.,}{{O'Connor}
  et~al.}{2016}]{OCoBecBla16}
{O'Connor} A.~P.,  et~al., 2016, \mn@doi [Phys. Rev. Lett.]
  {10.1103/PhysRevLett.116.113002}, 116, 113002

\bibitem[\protect\citeauthoryear{Sarre \& Whitham}{Sarre \&
  Whitham}{1988}]{SarWhi88}
Sarre P.~J.,  Whitham C.~J.,  1988, \mn@doi [Chem. Phys.]
  {10.1016/0301-0104(88)87067-8}, 124, 439

\bibitem[\protect\citeauthoryear{Sauer \& {\u{S}}pirko}{Sauer \&
  {\u{S}}pirko}{2013}]{SauSpi13}
Sauer S. P.~A.,  {\u{S}}pirko V.,  2013, \mn@doi [{J. Chem. Phys.}]
  {10.1063/1.4774374}, 138, 024315

\bibitem[\protect\citeauthoryear{Saxon \& Liu}{Saxon \& Liu}{1983}]{SaxLiu83}
Saxon R.~P.,  Liu B.,  1983, \mn@doi [{J. Chem. Phys.}] {10.1063/1.444873}, 78,
  1344

\bibitem[\protect\citeauthoryear{Saxon, Kirby  \& Liu}{Saxon
  et~al.}{1980}]{SaxKirLiu80}
Saxon R.~P.,  Kirby K.,   Liu B.,  1980, \mn@doi [{J. Chem. Phys.}]
  {10.1063/1.440323}, 73, 1873

\bibitem[\protect\citeauthoryear{Seleznev \& Khrustov}{Seleznev \&
  Khrustov}{2012}]{SelKhr12}
Seleznev A.~O.,  Khrustov V.~F.,  2012, \mn@doi [Russ. J. Phys. Chem. B]
  {10.1134/s1990793112060188}, 6, 681

\bibitem[\protect\citeauthoryear{Seleznev, Khrustov  \& Stepanov}{Seleznev
  et~al.}{2013}]{SelKhrSte13}
Seleznev A.~O.,  Khrustov V.~F.,   Stepanov N.~F.,  2013, \mn@doi [Chem. Phys.
  Lett.] {10.1016/j.cplett.2013.10.010}, 588, 253

\bibitem[\protect\citeauthoryear{{Smith}, {Liszt}  \& {Lutz}}{{Smith}
  et~al.}{1973}]{SmiLisLut73}
{Smith} W.~H.,  {Liszt} L.~S.,   {Lutz} B.~L.,  1973, \mn@doi [\apj]
  {10.1086/152209}, \href {http://adsabs.harvard.edu/abs/1973ApJ...183...69S}
  {183, 69}

\bibitem[\protect\citeauthoryear{{Solomon} \& {Klemperer}}{{Solomon} \&
  {Klemperer}}{1972}]{SolKle72}
{Solomon} P.~M.,  {Klemperer} W.,  1972, \mn@doi [\apj] {10.1086/151799}, \href
  {http://adsabs.harvard.edu/abs/1972ApJ...178..389S} {178, 389}

\bibitem[\protect\citeauthoryear{{Stancil}, {Babb}  \& {Dalgarno}}{{Stancil}
  et~al.}{1993}]{StaBabDal93}
{Stancil} P.~C.,  {Babb} J.~F.,   {Dalgarno} A.,  1993, \mn@doi [\apj]
  {10.1086/173113}, \href {http://adsabs.harvard.edu/abs/1993ApJ...414..672S}
  {414, 672}

\bibitem[\protect\citeauthoryear{Stancil et~al.,}{Stancil
  et~al.}{1998a}]{StaGuHav98}
Stancil P.~C.,  et~al., 1998a, \mn@doi [J. Phys. B: At. Mol. Opt. Phys.]
  {10.1088/0953-4075/31/16/017}, 31, 3647

\bibitem[\protect\citeauthoryear{Stancil et~al.,}{Stancil
  et~al.}{1998b}]{StaHavKrs98}
Stancil P.~C.,  et~al., 1998b, \mn@doi [\apj] {10.1086/305937}, 502, 1006

\bibitem[\protect\citeauthoryear{Stecher \& Williams}{Stecher \&
  Williams}{1974}]{SteWil74}
Stecher T.~P.,  Williams D.~A.,  1974, \mn@doi [\mnras]
  {10.1093/mnras/168.1.51P}, 168, 51P

\bibitem[\protect\citeauthoryear{{Talbi} \& {DeFrees}}{{Talbi} \&
  {DeFrees}}{1991}]{TalDeF91}
{Talbi} D.,  {DeFrees} D.~J.,  1991, \mn@doi [Chem. Phys. Lett.]
  {10.1016/0009-2614(91)90309-W}, \href
  {http://adsabs.harvard.edu/abs/1991CPL...179..165T} {179, 165}

\bibitem[\protect\citeauthoryear{Tielens}{Tielens}{2005}]{Tie05}
Tielens A.,  2005, The Physics and Chemistry of the Interstellar Medium.
Cambridge University Press, Cambridge

\bibitem[\protect\citeauthoryear{Vissapragada, Buzard, Miller, O'Connor, de
  Ruette, Urbain  \& Savin}{Vissapragada et~al.}{2016}]{VisBuzMil16}
Vissapragada S.,  Buzard C.~F.,  Miller K.~A.,  O'Connor A.~P.,  de Ruette N.,
  Urbain X.,   Savin D.~W.,  2016, \mn@doi [\apj] {10.3847/0004-637x/832/1/31},
  832, 31

\bibitem[\protect\citeauthoryear{Watson}{Watson}{2008}]{Watson2008}
Watson J. K.~G.,  2008, \mn@doi [{J. Molec. Spectrosc.}]
  {10.1016/j.jms.2008.04.014}, \textbf{253}, 5

\bibitem[\protect\citeauthoryear{Werner \& Knowles}{Werner \&
  Knowles}{1985}]{WerKno85}
Werner H.-J.,  Knowles P.~J.,  1985, \mn@doi [J. Chem. Phys.]
  {10.1063/1.448627}, 82, 5053

\bibitem[\protect\citeauthoryear{Werner, Knowles, Knizia, Manby, {Sch\"{u}tz}
  et~al.}{Werner et~al.}{2015}]{MOLPRO_brief}
Werner H.-J.,  Knowles P.~J.,  Knizia G.,  Manby F.~R.,  {Sch\"{u}tz} M.,
  et~al., 2015, \textsc{molpro}, version 2015.1, a package of \textit{ab
  initio} programs

\bibitem[\protect\citeauthoryear{Williams}{Williams}{1992}]{Wil92}
Williams D.~A.,  1992, \mn@doi [Planet. Space. Sci.]
  {10.1016/0032-0633(92)90125-8}, 40, 1683

\makeatother
\end{thebibliography}

\bsp	
\label{lastpage}
\end{document}